\documentclass[11pt]{article}

\usepackage{arxiv}

\usepackage{psfrag,epsf}
\usepackage{enumerate}
\usepackage[utf8]{inputenc} 
\usepackage[T1]{fontenc}    
\usepackage{hyperref}       
\usepackage{url}            
\usepackage{booktabs}       
\usepackage{amsfonts}       
\usepackage{nicefrac}       
\usepackage{microtype}      
\usepackage{float}          
\usepackage{natbib}         
\usepackage{bm}             
\usepackage{amssymb}        
\usepackage{amsmath}        
\usepackage{graphicx}       
\usepackage{comment}
\usepackage[font=footnotesize,labelfont=bf]{caption}
\usepackage{subcaption}
\usepackage{tikz}           
\usepackage[shortlabels]{enumitem}
\usepackage{array}          
\usepackage{lscape}         
\usepackage{dirtytalk}      
\usepackage{mathtools}      
\usepackage{tabularx}  
\usepackage{lmodern}
\usepackage{siunitx}
\usepackage{etoolbox}
\usepackage{physics}  
\usepackage{cleveref}             
\usepackage{doi}

\DeclareMathOperator{\logit}{logit}

\title{Joint model for longitudinal and spatio-temporal survival data}

\date{}

\author{ \href{https://orcid.org/0000-0002-3695-1436}{\includegraphics[scale=0.06]{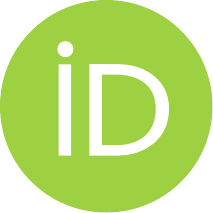}\hspace{1mm}Victor Medina-Olivares}
	\\
    Research Center Trustworthy Data Science and Security, UA Ruhr\\
    Department of Statistics, TU Dortmund, Dortmund, Germany\\
	\texttt{victor.medina@tu-dortmund.de}%
    \And
	\href{https://orcid.org/0000-0002-5833-2011}{\includegraphics[scale=0.06]{orcid.pdf}\hspace{1mm}Finn Lindgren}\\
    School of Mathematics, University of Edinburgh, Edinburgh, UK
    \And
	\href{https://orcid.org/0000-0002-0078-3151}{\includegraphics[scale=0.06]{orcid.pdf}\hspace{1mm}Raffaella Calabrese}\\
    Business School, University of Edinburgh, Edinburgh, UK\\
    \And
    \href{https://orcid.org/0000-0003-2198-4224}{\includegraphics[scale=0.06]{orcid.pdf}\hspace{1mm}Jonathan Crook}\\
    Business School, University of Edinburgh, Edinburgh, UK
}

\renewcommand{\shorttitle}{Spatio-Temporal Joint Model (STJM)}

\hypersetup{
pdftitle={Joint model for longitudinal and spatio-temporal survival data},
pdfsubject={stat.AP, q-fin.RM},
pdfauthor={Victor Medina-Olivares, Finn Lindgren, Raffaella Calabrese, Jonathan Crook},
pdfkeywords={Discrete time-to-event, Spatio-temporal frailties, Bayesian joint model, Credit risk management},
}

\begin{document}

\maketitle

\begin{abstract}
In credit risk analysis, survival models with fixed and time-varying covariates are widely used to predict a borrower's time-to-event. When the time-varying drivers are endogenous, modelling jointly the evolution of the survival time and the endogenous covariates is the most appropriate approach, also known as the joint model for longitudinal and survival data. In addition to the temporal component, credit risk models can be enhanced when including borrowers' geographical information by considering spatial clustering and its variation over time. We propose the Spatio-Temporal Joint Model (STJM) to capture spatial and temporal effects and their interaction. This Bayesian hierarchical joint model reckons the survival effect of unobserved heterogeneity among borrowers located in the same region at a particular time. To estimate the STJM model for large datasets, we consider the Integrated Nested Laplace Approximation (INLA) methodology. We apply the STJM to predict the time to full prepayment on a large dataset of 57,258 US mortgage borrowers with more than 2.5 million observations. Empirical results indicate that including spatial effects consistently improves the performance of the joint model. However, the gains are less definitive when we additionally include spatio-temporal interactions.
\end{abstract}

\keywords{Discrete time-to-event \and Spatio-temporal frailties \and Bayesian joint model \and Credit risk management}

\section{Introduction}

Lenders build mathematical models to predict credit events like defaults and full prepayments. Survival approaches are popular in this regard, as they facilitate the inclusion of fixed and time-varying covariates (TVCs), handle censored data, and allow prediction for different time horizons \citep[see][]{thomas2017credit}. Recently, a new methodology has been introduced in this context, known as joint modelling of longitudinal and survival data \citep[joint models hereafter,][]{tsiatis2004joint, rizopoulos2012joint}. Initially developed in medical research, this methodology offers two attractive features compared to the standard survival credit risk approaches. First, when TVCs are endogenous, as typically in the case of performing covariates, joint models provide a robust statistical procedure to handle the mutual evolution of the survival process and endogenous TVCs. Second, by jointly modelling survival and endogenous TVCs, we encounter a natural prediction framework that does not rely on lagged values or exogeneity assumptions about TVCs, as is commonly done otherwise \citep{crook2010time}. In addition, recent studies show that joint models show better prediction performance than the survival counterparts \citep{hu2019joint,medina2023joint_auto,medina2023joint_inla}.

A joint model typically comprises two sub-models: one for the survival process and another for the endogenous TVC, also referred to as the longitudinal outcome. These sub-models are connected through a latent structure, often characterised by the inclusion of random effects. In this work, we considered adaptable predictor representations within the survival process, aiming to incorporate geographical information related to borrowers. This allows us to account for spatial clustering and its variation over time, factors that hold the potential to enhance the predictive capabilities of credit risk models, as shown previously \citep{goodstein2017contagion,gupta2019foreclosure,calabrese2020spatial,calabrese2023contagion}. 

We make four contributions to the literature: 
\begin{enumerate}[wide, labelwidth=!, labelindent=0pt]
    \item We propose a Bayesian hierarchical joint model in discrete time, featuring a flexible baseline hazard that accommodates spatial and spatio-temporal interactions. This approach captures the survival impact of unobserved spatial variables among borrowers and enables us to leverage information from neighbouring areas. We term this model the Spatio-Temporal Joint Model (STJM).
    \item To handle large datasets such as those seen in credit risk analysis, we adopt the INLA methodology \citep{rue2009approximate} to estimate the STJM. This allows us to estimate the model on a dataset with more than 2.5 million observations. To our knowledge, this dataset is the largest used in the joint model literature when writing this work.
    \item To compare different model specifications, we introduce a novel approximation method of the \emph{cross-validated Dynamic Conditional Likelihood} \citep[cvDCL, see][]{rizopoulos2016personalized}\footnote{The cvDCL is a cross-entropy estimate of the cross-validatory posterior predictive conditional density.}, leveraging pre-computed quantities from the INLA methodology for efficient estimation.
    \item We apply the STJM to predict full prepayment events in US mortgage loans and demonstrate that including spatial components consistently improves the performance of the joint model across all evaluation time points. Nevertheless, our empirical analysis reveals that the performance improvements are less significant when we incorporate spatio-temporal interactions in addition to the main effects.
\end{enumerate}

Previous studies show that spatial contagion plays a significant role in credit risk analysis on mortgage loans. \citet{goodstein2017contagion} obtain that surrounding areas have a relevant impact on strategic mortgage defaults in addition to borrowers' characteristics.  Strategic defaults occur when borrowers choose to default because the economic benefits outweigh the costs, unlike borrowers who default because they have an unexpected net income shock. Moreover, \citet{guiso2013determinants,towe2013contagion} find strong evidence that social interactions among neighbours influence the propensity for strategic defaults. There are different reasons for spatial contagion on mortgage defaults. One primary factor is the reduction of property values in a neighbourhood, which in turn is highly spatially correlated \citep{gelfand1998spatio, iversen2001spatially}. Neighbourhood characteristics, such as increasing crime rates, vandalisation, or legislative changes, can negatively impact a property value \citep{pence2006foreclosing}. Additionally, an increased number of defaults in certain areas can lead banks to limit credit options in these neighbours, such as renegotiations. Spatial contagion can also manifest in credit events beyond defaults. For example, \citet{gupta2019foreclosure} discover a significant spatial dependence in early repayment activity for mortgage loans due to similar reasons highlighted before for mortgage defaults. Decreased property values in a given area could affect borrowers' inclination to seek new refinancing options. At the same time, banks might reduce credit extensions or renegotiation if they anticipate a drop in property prices or other foreclosure externalities.

Regarding survival models for predicting mortgage defaults, \citet{calabrese2020spatial} is the first work to include spatial contagion. They incorporate time and spatial-varying coefficients in a survival model that predicts time to default in UK mortgage loans, showing better accuracy than relevant benchmarks. However, they do not account for possible endogeneity in the TVCs included in the model and lack a predictive framework for their future trajectories, as offered by the joint model approach.

Concerning the literature on joint models with spatial dependence, \citet{zhou2008joint} introduce a joint model in continuous time to handle two related time-to-event outcomes. They assume a Weibull baseline distribution with spatially correlated frailties. Additionally, \citet{ratcliffe2004joint} incorporate spatial clustering as univariate independent random effects. Building on the same principles, \citet{martins2016bayesian} propose a joint model with spatial random effects to analyse AIDS data in Brazil. In that work, they adopt an intrinsic conditional autoregressive model (ICAR, see \citet{besag1991bayesian}) as a prior distribution for the unobserved spatial effects, which aligns with our approach. However, these papers do not explore a unified joint model encompassing discrete survival data, spatio-temporal interactions, and do not demonstrate scalability with large datasets.

The manuscript is structured as follows. In Section \ref{sec:meth}, we introduce the STJM, its estimation procedure, and the Bayesian model selection. In Section \ref{sec:app}, we present and compare the empirical results of different joint models to predict the time to full prepayment event on US mortgages. Section \ref{sec:con} concludes.  

\section{Methodology} \label{sec:meth}
\subsection{Spatio-Temporal Joint Model (STJM)}\label{subsec:model}

Consider a total of $N$ mortgage loans with their associated properties distributed over $A$ areas. Each area, indexed by $a = 1,\ldots,A$, has a total of $N_a$ properties, i.e.\ $\sum_{a=1}^A N_a=N$. For each mortgage loan $i$ ($i=1,\ldots,N$), we are provided with the following information: the location $a_i\in \{1,\ldots,A\}$; the loan origination date $t^{0}{i}$; an event indicator $\delta_{i}$, which takes the value 1 if a full prepayment occurs and 0 otherwise; and the time elapsed from the loan origination to its last recorded observation time $t_{i}\le T$. Here, $T$ represents the duration of the study.
We assume that at time $t_{i}$, either a full prepayment has occurred ($\delta_{i}=1$), or the observation is right-censored ($\delta_{i}=0$), i.e.\ we observe loan $i$ until time $t_{i}$ but not beyond that \citep{PaulD.Allison1982}.

Additionally, we are provided with a vector of time-fixed covariates $\bm{z}_{i}$ and a loan-specific covariate collected regularly at multiple points in time (in our case, every month). We represent this time-varying covariate as $y_{i,s}$, where $s = 1,\ldots,t_{i}$. The $y_{i,s}$ values correspond to the longitudinal outcome within the framework of our joint model. We use lowercase to distinguish the realisations of random variables.

We aim to understand the relationship of these data in jointly modelling the time to event $T_{i}$ and the longitudinal outcome $Y_{i,s}$ up to a given endpoint for the $i$-th loan associated with area $a_i$. The following describes the proposed approach for the longitudinal and survival processes.  

\subsubsection*{Longitudinal process} 

Assume the longitudinal outcome $Y_{i,s}$ follows a mixed-effect model \citep{laird1982random}, where the predictor $\eta_{Yi,s}$ is composed of fixed effects $\bm{q}^{\intercal}_{i,s}\bm{\beta}_1$ and random effects $\bm{d}_{i,s}^{\intercal} \bm{U}_{i}$. Here, $\bm{\beta}_1$ is a vector of coefficients associated with the covariates $\bm{q}_{i,s}$, and $\bm{d}_{i,s}$ is the design vector corresponding to the random effects $\bm{U}_{i}$ of dimension $r$. Specifically,
\begin{equation}\label{eq:long}
    \begin{split}
    Y_{i,s}|\eta_{Yi,s}, \tau_Y & \sim N(\eta_{Yi,s},\tau_Y^{-1})\\
    \eta_{Yi,s} &= \bm{q}^{\intercal}_{i,s}\bm{\beta}_1 + \bm{d}_{i,s}^\intercal \bm{U}_{i}\\
    \bm{U}_{i}|Q_{\bm{U}} &\sim N_r(\bm{0},Q_{\bm{U}}^{-1}),
    \end{split}
\end{equation} 
where $\tau_Y$ is the precision parameter of the error terms. We assume that $\bm{U}_{i}$ are mutually independent among mortgage loans and distributed as a zero-mean multivariate Gaussian distribution with $r\times r$ precision matrix $Q_{\bm{U}}$. Given the random effects, we consider that observations within each loan are conditionally independent. Therefore, the random effects account for the correlation between these different observations.

\subsubsection{Survival process}
Following the discrete-time survival formulation presented in \citet{PaulD.Allison1982}, we represent the random variable $T_{i}$ using a sequence of binary random variables $X_{i,s}$. These variables take the value 1 if the loan $i$ is fully prepaid at time $s = t_i$ after origination and 0 otherwise. In the case of censored loans, the sequence will consist entirely of zeros. Conversely, for fully prepaid loans, the sequence will be composed of zeros, except for the last observation, which will take the value 1. To relate $X_{i,s}$ with the predictor $\eta_{Xi,s}$, we use a logit link function, which can be expressed as follows
\begin{equation}\label{eq:event}
    \begin{split}
    X_{i,s}|\eta_{Xi,s} &\sim \text{Bernoulli}(\logit^{-1}( \eta_{Xi,s})) \\
    \eta_{Xi,s} &= \nu_{a_i, s} + \bm{z}^{\intercal}_{i} \bm{\beta}_2 + \lambda (\bm{d}_{i,s}^\intercal \bm{U}_{i}) ,
    \end{split}
\end{equation} 

where $\nu_{a_i, s}$ represents the baseline risk, which varies across both time and space, covering the entire discrete domain of $a_i \in \{1,\ldots,A\}$ and $s \in \{1,\ldots,T\}$. The vector $\bm{\beta}_2$ contains coefficients associated with the covariates $\bm{z}_{i}$. The parameter $\lambda$ indicates the association between the survival time and the random effects $\bm{d}_{i,s}^\intercal \bm{U}_{i}$. Consequently, the random effects play a crucial role in both the longitudinal and survival processes. In the longitudinal process (Equation \ref{eq:long}), they account for the correlation between repeated measurements. In the survival process (Equation \ref{eq:event}), along with $\lambda$, they account for the degree of association with the longitudinal outcome.

Following \citet{chang2013spatial}, who present an additive decomposition of spatio-temporal effects for a survival model, we take a similar approach in our joint model. Specifically, we express $\nu_{a,s} = \nu_0 + v_s + u_a + \delta_{a,s}$. Here, $\nu_0$ represents the overall average, $v_s$ denotes the temporal main effect, $u_a$ accounts for the spatial main effect, and $\delta_{a,s}$ captures the spatio-temporal interaction. In the following sections, we describe each term, namely $v_s$, $u_a$, and $\delta_{a,s}$.

\paragraph*{Temporal main effects ($v_s$)} The vector of temporal effects is denoted as $\bm{v} = (v_1,\ldots,v_T)^\intercal$. We model these effects using a second-order random walk model \citep[see][]{lindgren2008second}, which is characterised by the following joint density
\begin{equation} \label{eq:rw2} 
 \begin{split}
    \bm{v} | \tau_{v} &\propto  \exp\left(-\frac{\tau_v}{2} \sum_{s = 3}^{T}(v_s-2v_{s-1}+v_{s-2})^2\right)\\
    & = \exp\left(-\frac{\tau_v}{2} \bm{v}^\intercal R_v \bm{v} \right),
 \end{split}
\end{equation}
where $\tau_{v}$ is a precision parameter and the $T \times T$ matrix $R_v$ is the so-called \emph{structure matrix} \citep{rue2005gaussian} defined as (the zeros are not shown)
\begin{equation*}
R_v = \begin{pmatrix}
1 & -2 & 1 & & & &  \\
-2 & 5 & -4 & 1 & & & &   \\
1 & -4 & 6 & -4 & 1 & & &  \\
& \ddots & \ddots & \ddots & \ddots & \ddots & &  \\
 & & 1 & -4 & 6 & -4 & 1 \\
 & & & 1 & -4 & 5 & -2\\
 & & & & 1 & -2 & 1 
\end{pmatrix}.
\end{equation*}

\paragraph*{Spatial main effects ($u_a$)} Regarding the spatial effects $\bm{u} = (u_1,\ldots,u_A)^\intercal$, we adopt an intrinsic conditional autoregressive model \citep[ICAR, see][]{besag1991bayesian}. This model suggests that neighbouring areas may exhibit similar repayment behaviour (as shown in studies like \citet{calabrese2020spatial,calabrese2023contagion} for default prediction). The joint density for the ICAR model is expressed as
\begin{equation} \label{eq:besag}
    \bm{u} | \tau_{u} \propto \exp\left(-\frac{\tau_u}{2} \sum_{a\sim a'} (u_a-u_{a'})^2\right),
\end{equation}

where $\tau_u$ serves as a precision parameter, and the notation $a\sim a'$ indicates that the two areas are neighbours. The definition of \say{neighbour} can be varied and depends on the specific application \citep{freni2018note}. In this study, we adhere to the standard definition, where two areas are considered neighbours if they share a common border. Alternatively, one can define connected areas based on the distance between their centroids \citep[e.g.][]{goodstein2017contagion,medina2022spatial}. However, exploring the different ways to define neighbours is beyond the scope of this work, and interested readers are directed to \citet[][Ch. 4]{banerjee2014hierarchical} for further discussion on this topic.

For this specification, the corresponding elements of the $A\times A$ structure matrix $R_u$ of Equation \ref{eq:besag} are
\begin{equation*}
    (R_u)_{aa'} = 
     \begin{cases}
       m_a \quad a=a'\\
       -1 \quad a \sim a'\\
       0  \quad\text{otherwise},
     \end{cases}
\end{equation*}
where $m_a$ is the number of neighbours of area $a$. The full conditional density of the ICAR model enables a more accessible interpretation, given by
\begin{equation*} 
    u_a|\bm{u}_{-a}, \tau_u \sim N\left (\frac{1}{m_a}\sum_{a': a\sim a'} u_{a'}, \frac{1}{\tau_u m_a}\right ) ,
\end{equation*}
where $\bm{u}_{-a}$ represents the set of spatial effects excluding the area $a$. Therefore, $u_a$ has a local mean of $\sum_{a': a\sim a'} u_{a'}/m_a$, which corresponds to the average value of spatial effects from the neighbouring areas, and its variance is inversely related to the number of neighbours, $m_a$. Consequently, the presence of more neighbours results in greater certainty regarding the effect.

\paragraph*{Spatio-temporal interactions ($\delta_{a,s}$)} 
To model the spatio-temporal interactions $\bm{\delta} = (\delta_{11},\ldots,\delta_{A1},\ldots,\allowbreak\delta_{1T},\ldots,\delta_{AT})^\intercal$, we adopt the approach presented in \citet{clayton1996generalized} and further elaborated in \citet{knorr2000bayesian}. In this approach, the structure matrix $R_\delta$ can be obtained as the Kronecker product of the structure matrices from the temporal and spatial main effects, i.e.\ $R_\delta = R_v \otimes R_u$. As a result, the corresponding joint density is given by \citet{schrodle2011spatio}
\begin{equation} \label{eq:interiv}
    \bm{\delta} | \tau_{\delta} \propto \exp\Big(-\frac{\tau_\delta}{2} \sum_{s = 3}^{T} \sum_{a\sim a'} \big[(\delta_{a,s} - 2\delta_{a,s-1} +\delta_{a,s-2})-  
    (\delta_{a',s-2} - 2\delta_{a',s-1} +\delta_{a',s} ) \big]^2\Big),
\end{equation}
where $\tau_\delta$ is the corresponding precision parameter.

In the spatial literature, it is widely recognised that structured additive predictors formed by Equations \ref{eq:rw2}, \ref{eq:besag}, and \ref{eq:interiv} can lead to identifiability issues \citep[see, e.g.][]{knorr2000bayesian, goicoa2018spatio}. To ensure appropriate identifiability, we need to impose constraints on the random effects $\bm{v}$, $\bm{u}$ and $\bm{\delta}$. In this regard, we follow the approach of \citet{goicoa2018spatio}, who employ reparametrisations using spectral decomposition on the structure matrices $R_v$, $R_u$ and $R_\delta$. These reparametrisations conduct to the following constraints: $\sum_{s=1}^T v_s=0$, $\sum_{a=1}^A u_a=0$, $\sum_{s=1}^T\delta_{a,s}=0$ for $a=1,\ldots,A$ and $\sum_{a=1}^A\delta_{a,s}=0$ for $s=1,\ldots,T$.

\subsection{Estimation} \label{subsec:est}

From Equations \ref{eq:long} and \ref{eq:event}, we know that the random effects $\mathbf{U}i$ are shared between both the longitudinal and survival processes. The joint model approach assumes that these two processes are conditionally independent given the random effects \citep{wulfsohn1997joint, henderson2000joint, tsiatis2004joint}. Therefore, the joint distribution of the observed values $\bm{y}_i = (y_{i1},\ldots,y_{i,t_i})^\intercal$ and $\bm{x}_i= (x_{i1},\ldots,x_{i,t_i})^\intercal$ for loan $i$ conditional on the random effects is 
\begin{equation} \label{eq:cond_p}
    p(\bm{y}_i, \bm{x}_i|\bm{U}_i,\Theta) = \prod_{s=1}^{t_i} p(y_{i,s}|\bm{U}_i,\Theta)p(x_{i,s}|\bm{U}_i,\Theta),
\end{equation}
where $\Theta$ denotes the vector of parameters included in both processes. It follows from Equation \ref{eq:long} that
\begin{equation*}
 \begin{split}
    p(y_{i,s}|\bm{U}_i,\Theta) &= \left(\frac{\tau_Y}{2\pi}\right)^{1/2}\exp \left(-\frac{\tau_Y(y_{i,s}-\eta_{Yi,s})^2}{2} \right)\\
    &=\left(\frac{\tau_Y}{2\pi}\right)^{1/2}\exp \left(-\frac{\tau_Y(y_{i,s}-\bm{q}^{\intercal}_{i,s}\bm{\beta}_1 - \bm{d}_{i,s}^\intercal \bm{U}_{i})^2}{2} \right),
 \end{split}
\end{equation*}
and from Equation \ref{eq:event}
\begin{equation*}
 \begin{split}
    p(x_{i,s}|\bm{U}_i,\Theta) &= [\logit^{-1}( \eta_{Xi,s})]^{x_{i,s}}[1-\logit^{-1}( \eta_{Xi,s})]^{1-x_{i,s}}\\
    &=[\logit^{-1}( \nu_{a_i, s} + \bm{z}^{\intercal}_{i} \bm{\beta}_2 + \lambda (\bm{d}_{i,s}^\intercal \bm{U}_{i}))]^{x_{i,s}}\\
    &\quad \times[1-\logit^{-1}( \nu_{a_i, s} + \bm{z}^{\intercal}_{i} \bm{\beta}_2 + \lambda (\bm{d}_{i,s}^\intercal \bm{U}_{i}))]^{1-x_{i,s}}.
 \end{split}
\end{equation*}
Hence, the contribution of the $i$-th loan to the observation density is
\begin{equation}
    \begin{split}
       p(\bm{y}_i, \bm{x}_i | \Theta) &= \int p(\bm{y}_i, \bm{x}_i|\bm{U}_i,\Theta)p(\bm{U}_i|\Theta) \mathrm{d}\bm{U}_i\\
       &= \int\prod_{s=1}^{t_i} p(y_{i,s}|\bm{U}_i,\Theta)p(x_{i,s}|\bm{U}_i,\Theta)p(\bm{U}_i|\Theta) \mathrm{d}\bm{U}_i,
    \end{split}
\end{equation}
where $p(\bm{U}_i|\Theta)$ is as zero-mean multivariate Gaussian with precision matrix $Q_{\bm{U}}$ (Section \ref{subsec:model}), i.e.\ $p(\bm{U}_i|\Theta) = (2\pi)^{-r/2}|Q_{\bm{U}}|^{1/2}\exp\left (-\bm{U}_i^\intercal Q_{\bm{U}}\bm{U}_i/2 \right )$. 

Denote the complete set of observation variables as $\mathcal{D} = \{\bm{y}_{i},\bm{x}_{i} : i = 1,\ldots,N\}$. The joint posterior distribution follows $p(\Theta|\mathcal{D}) \propto p(\mathcal{D}|\Theta)p(\Theta)$, where $p(\mathcal{D}|\Theta)= \prod_i^N p(\bm{y}_i, \bm{x}_i | \Theta)$ is the overall observation density and $p(\Theta)$ the joint prior. 

Theoretically, we could estimate this model specification using simulation-based schemes, as demonstrated in \citet{medina2023joint_auto} for the joint model with autoregressive terms. However, it should be noted that this strategy can be computationally expensive and may even become infeasible for applications dealing with large datasets. To address these computational challenges and in line with the estimation approach followed in \citet{medina2023joint_inla}, we propose employing the INLA methodology \citep{rue2009approximate}.

INLA offers accurate estimations of the posterior at a considerably lower computational cost and is readily accessible through the \texttt{R-INLA} software package for R (\url{https://www.r-inla.org/}). This methodology is particularly suitable for models belonging to the class of latent Gaussian models (LGM), which is a flexible and widely used class of models. For instance, many structured Bayesian additive models fall under this category \citep[see][]{fahrmeir1994multivariate, gelman2013bayesian}. Importantly, the STJM also belongs to this class of models, as shown next.

We define the latent field $\bm{\mu} = (\bm{\eta}_Y, \bm{\eta}_X, \bm{U},\bm{\beta}_1,\allowbreak\bm{\beta}_2, \nu_0, \bm{v}, \bm{u}, \bm{\delta})$, which comprises the set of unobserved variables in the STJM. The terms $\bm{\eta}_Y$ and $\bm{\eta}_X$ correspond to the predictors described in Equations \ref{eq:long} and \ref{eq:event}, respectively, each having $\sum_{i}^{N} t_i$ elements. As the remaining elements are latent variables, we refer to $\bm{\mu}$ as a latent field. Additionally, since we assume that $\bm{\mu}$ follows a zero-mean multivariate Gaussian distribution, it is called a latent Gaussian field \citep{rue2005gaussian}.

Expressly, we assume that the coefficients $\bm{\beta}_1$, $\bm{\beta}_2$, and $\nu_0$ follow a zero-mean Gaussian distribution with a precision matrix $\tau_{f} \bm{I}$, where $\bm{I}$ is the identity matrix of the corresponding dimension, and $\tau_f$ is a precision parameter. Typically, $\tau_f$ is set as a fixed value close to zero in the model, resulting in a large prior variance.

As described in Section \ref{subsec:model}, the random effects $\bm{U}_i|Q_{\bm{U}} \sim N(\bm{0},Q_{\bm{U}}^{-1})$, and the terms $\bm{v}$, $\bm{u}$, and $\bm{\delta}$ have priors with Gaussian kernels (see Equations \ref{eq:rw2}, \ref{eq:besag}, and \ref{eq:interiv}, respectively). Consequently, the precision matrix of the latent Gaussian field $\bm{\mu}$, which includes all the individual precision matrices, is denoted as $Q(\bm{\theta}_1)$, where $\bm{\theta}_1$ is the corresponding set of hyperparameters. In our case, $\bm{\theta}_1=(\tau_f, Q_{\bm{U}}, \lambda, \tau_v, \tau_u, \tau_{\delta})$.

Despite the potentially large dimension of the matrix $Q(\bm{\theta}_1)$, INLA benefits from computation efficiency due to the sparsity of this matrix \citep{rue2009approximate}.

Moreover, let $\bm{\theta}_2$ represent the set of hyperparameters directly affecting the observation density, which, in our case, includes only the precision parameter $\tau_Y$. We can restate Equation \ref{eq:cond_p} using the INLA notation as $p(\bm{y}_i,\bm{x}_i|\bm{U}_i,\Theta)=\prod_{s=1}^{t_i} p(\mathcal{D}_{i(s)} |\mu_{i(s)}, \bm{\theta_2})$. This reformulation allows us to express the overall observation density as $p(\mathcal{D}|\bm{\mu},\bm{\theta}_2) = \prod_i^N \prod_{s=1}^{t_i} p(\mathcal{D}_{i(s)} |\mu_{i(s)}, \bm{\theta_2})$, which can be further simplified to $p(\mathcal{D}|\bm{\mu},\bm{\theta}_2) = \prod_{j \in \mathcal{J}} p(\mathcal{D}_{j} |\mu_{j}, \bm{\theta_2})$ by changing the corresponding indexes. This last expression demonstrates, in line with the requirements of the INLA methodology, that the observation density is conditionally independent.

By denoting the complete set of hyperparameters as $\bm{\theta} = (\bm{\theta}_1, \bm{\theta}_2)$, we recover the same formulation described in \citet{rue2009approximate}. This confirms that the STJM falls within the class of latent Gaussian models, making it suitable for the INLA estimation. More details on how posterior marginals and their corresponding numerical integrations with INLA are computed can be found in \citet{rue2009approximate,medina2023joint_inla}.

\subsection{Bayesian model selection with INLA} \label{subsec:inla_cvdcl}
We aim to select the model that best predicts the full prepayment event of loan $i$ at a given time point $t$, assuming that the loan has not been repaid up to that time. To compare the predictive performance of different models based on the collected observations, we adopt the methodology proposed by \citet{rizopoulos2016personalized}, which we extend to both the INLA estimation procedure and the STJM formulation.

The authors suggest choosing the model that minimises the cross-entropy of the survival outcome's cross-validatory posterior predictive conditional density. Concretely, for each model $M_k \in \{M_1, \ldots,M_K\}$ and at time $t$, we seek to estimate $p(T_{i} | T_{i}>t,\bm{y}_{i}(t),\mathcal{D}_{-i}, M_k)$. Here, $\bm{y}{i}(t)$ represents the historical observations of the longitudinal outcome of loan $i$ up to time $t$, i.e.\ $\bm{y}_{i}(t) = \{y_{i,s}:s\le t\}$, and $\mathcal{D}_{-i}$ denotes the data excluding loan $i$.

The best model, denoted as $M_{\tilde k}$, where $\tilde k \in \{1,\ldots,K\}$, is determined by minimising the cross-entropy $E(-\log\{p(T_{i} | T_{i}>t, \bm{y}_{i}(t),\mathcal{D}_{-i}, M_{\tilde k} ) \})$. The expectation is taken with respect to the model that generated the data, which might not necessarily be one of the $K$ models considered in practice.

To account for the censored cases, \citet{rizopoulos2016personalized} propose to use the available information and termed this estimate as the \emph{cross-validated Dynamic Conditional Likelihood} (cvDCL) defined as\footnote{Although we do not explicitly indicate the model $k$, it is implicitly assumed in the conditioning.}
\begin{equation} \label{eq:cvDCL}
    \text{cvDCL}(t) = \frac{1}{N_t}\sum_{i = 1}^N-I(T_i>t)\log\{p(T_i,\delta_i | T_i>t, \bm{y}_{i}(t),\mathcal{D}_{-i} ) \},
\end{equation}
where $N_t$ is the number of loans at risk at time $t$, i.e.\ $N_t = \sum_{i=1}^N I(T_i>t)$.

Once the model is estimated, INLA allows generating samples from the approximated posterior density. We propose using this feature to calculate the expression in Equation \ref{eq:cvDCL} through Monte Carlo integration, as shown below. 

First, note that
\begin{equation*}
    \frac{p(\bm{\theta}|T_i, \delta_i, T_i>t,\bm{y}_{i}(t), \mathcal{D}_{-i})p(T_i, \delta_i| T_i>t,\bm{y}_{i}(t), \mathcal{D}_{-i}) }{p(T_i, \delta_i| T_i>t,\bm{y}_{i}(t), \mathcal{D}_{-i}, \bm{\theta}) } = p(\bm{\theta}| T_i>t,\bm{y}_{i}(t), \mathcal{D}_{-i}),
\end{equation*}
and integration of this last expression with respect to $\bm{\theta}$ leads to
\begin{equation} \label{eq:cvDCL_exp1}
    \begin{split}
    p(T_i, \delta_i| T_i>t,\bm{y}_{i}(t), \mathcal{D}_{-i})^{-1} &= \int \frac{p(\bm{\theta}|T_i, \delta_i, T_i>t,\bm{y}_{i}(t), \mathcal{D}_{-i}) }{p(T_i, \delta_i| T_i>t,\bm{y}_{i}(t), \mathcal{D}_{-i}, \bm{\theta}) }\mathrm{d}\bm{\theta}\\
     &\approx \int \frac{p(\bm{\theta}|\mathcal{D})}{p(T_i,\delta_i | T_i>t, \bm{y}_{i}(t),\mathcal{D}_{-i},\bm{\theta} )}  \mathrm{d}\bm{\theta} \\ 
     &\approx \sum_w \frac{\hat{p}(\bm{\theta}_w|\mathcal{D})}{\hat{p}(T_i,\delta_i | T_i>t, \bm{y}_{i}(t),\mathcal{D}_{-i},\bm{\theta}_w)} \Delta_w .
     \end{split}
\end{equation}
The integration grid $\{\bm{\theta}_w,\Delta_w\}$ of $\bm{\theta}$ is constructed by INLA when estimating the model, where $\Delta_w$ represents the integration weights.  

Moreover, note that the denominator $p(T_i,\delta_i | T_i>t, \bm{y}_{i}(t),\mathcal{D}_{-i},\bm{\theta}_w)$ follows
\begin{displaymath}
    \frac{p(\bm{\mu}|T_i, \delta_i, T_i>t,\bm{y}_{i}(t), \mathcal{D}_{-i}, \bm{\theta}_w )p(T_i, \delta_i| T_i>t,\bm{y}_{i}(t), \mathcal{D}_{-i}, \bm{\theta}_w ) }{p(T_i, \delta_i| T_i>t,\bm{y}_{i}(t),  \bm{\theta}_w,\bm{\mu} ) }
     = p(\bm{\mu}| T_i>t,\bm{y}_{i}(t), \mathcal{D}_{-i},\bm{\theta}_w),
\end{displaymath}
where $\bm{\mu}$ is the latent field described in Section \ref{subsec:est}. Thus, integrating this last expression with respect to $\bm{\mu}$ gives
\begin{align*} 
    p(T_i, \delta_i| T_i>t,\bm{y}_{i}(t), &\mathcal{D}_{-i}, \bm{\theta}_w )^{-1} =  \int \frac{p(\bm{\mu}|T_i, \delta_i, T_i>t,\bm{y}_{i}(t), \mathcal{D}_{-i}, \bm{\theta}_w ) }{p(T_i, \delta_i| T_i>t,\bm{y}_{i}(t), \bm{\theta}_w,\bm{\mu} ) } \mathrm{d}\bm{\mu}\\
    &=  \int \frac{p(\bm{U}_i,\bm{\mu}_{-\bm{U}_i}|T_i, \delta_i, T_i>t,\bm{y}_{i}(t), \mathcal{D}_{-i}, \bm{\theta}_w ) }{p(T_i, \delta_i| T_i>t,\bm{y}_{i}(t),  \bm{\theta}_w,\bm{U}_i,\bm{\mu}_{-\bm{U}_i} ) } \mathrm{d}\bm{\mu}_{-\bm{U}_i} \mathrm{d}\bm{U}_i \\
    & \approx  \int \frac{p(\bm{U}_i| T_i>t,\bm{y}_{i}(t), \bm{\theta}_w,\bm{\mu}_{-\bm{U}_i} ) p(\bm{\mu}_{-\bm{U}_i}|\mathcal{D}, \bm{\theta}_w ) }{p(T_i, \delta_i| T_i>t,\bm{y}_{i}(t),  \bm{\theta}_w,\bm{U}_i,\bm{\mu}_{-\bm{U}_i} ) } \mathrm{d}\bm{\mu}_{-\bm{U}_i} \mathrm{d}\bm{U}_i.
\end{align*}
We use the notation $\bm{\mu}= (\bm{U}_i,\bm{\mu}_{-\bm{U}_i})^\intercal$ to separate the random effects $\bm{U}_i$ that strictly depend on the loan $i$ from the rest of the parameters $\bm{\mu}_{-\bm{U}_i}$.

Let $\bm{\mu}_{-\bm{U}_i}^{(r,w)}$ denotes the $r$th realisation of the approximated posterior sample with $r = 1,\ldots,R$, then $p(T_i, \delta_i| T_i>t,\bm{y}_{i}(t), \mathcal{D}_{-i})^{-1}$ from Equation \ref{eq:cvDCL_exp1}, can be estimated as
\begin{align*}
    p(T_i, &\delta_i| T_i>t,\bm{y}_{i}(t), \mathcal{D}_{-i})^{-1} \approx \sum_w \frac{\hat{p}(\bm{\theta}_w|\mathcal{D})}{\hat{p}(T_i,\delta_i | T_i>t, \bm{y}_{i}(t),\mathcal{D}_{-i},\bm{\theta}_w)} \Delta_w \\
    & \approx  \sum_w \hat{p}(\bm{\theta}_w|\mathcal{D}) \Delta_w \int \frac{p(\bm{U}_i| T_i>t,\bm{y}_{i}(t), \bm{\theta}_w,\bm{\mu}_{-\bm{U}_i} ) p(\bm{\mu}_{-\bm{U}_i}|\mathcal{D}, \bm{\theta}_w ) }{p(T_i, \delta_i| T_i>t,\bm{y}_{i}(t),  \bm{\theta}_w,\bm{U}_i,\bm{\mu}_{-\bm{U}_i} ) } \mathrm{d}\bm{\mu}_{-\bm{U}_i} \mathrm{d}\bm{U}_i \\
    & \approx \sum_w \hat{p}(\bm{\theta}_w|\mathcal{D}) \Delta_w \left[\frac{1}{R}\sum_r \int \frac{p(\bm{U}_i| T_i>t,\bm{y}_{i}(t), \bm{\theta}_w,\bm{\mu}_{-\bm{U}_i}^{(r,w)} ) }{p(T_i, \delta_i| T_i>t,\bm{y}_{i}(t),  \bm{\theta}_w,\bm{U}_i,\bm{\mu}_{-\bm{U}_i}^{(r,w)} ) } \mathrm{d}\bm{U}_i \right].
\end{align*}

Furthermore, the integral can be calculated, for instance, with empirical Bayes or the Laplace method \citep{tierney1986accurate}. Whichever method is used to calculate the integral, denote this term as $h_i(\bm{\theta}_w,\bm{\mu}_{-\bm{U}_i}^{(r,w)} | t)$. Hence, $\text{cvDCL}(t)$ can be estimated as
\begin{equation} \label{eq:est_cvDCL_inla}
    \widehat{\text{cvDCL}}(t)^{INLA} = \frac{1}{N_t}\sum_{i = 1}^N I(T_i>t) \times \log\left\{ \sum_w \hat{p}(\bm{\theta}_w|\mathcal{D}) \Delta_w \left[\frac{1}{R}\sum_r h_i(\bm{\theta}_w,\bm{\mu}_{-\bm{U}_i}^{(r,w)} | t ) \right] \right\}.
\end{equation}

To get an estimate of the Monte Carlo variance of Equation \ref{eq:est_cvDCL_inla}, we use what is known as the \emph{Delta method} \citep{ver2012invented}. This method approximates a function of random variables using a Taylor series expansion around the means. In our case, we can identify the random variables as $h_{iwr|t}=h_i(\bm{\theta}_w,\bm{\mu}_{-\bm{U}_i}^{(r,w)} | t )$ which are independent for all the loans $i$, the integration points $w$ and the realisations $r$. Denote $m_{iw|t} = \mathrm{E}(h_{iwr|t})$ and $\sigma_{iw|t}^2=\mathrm{Var}(h_{iwr|t})$ and their estimations, respectively, as $\hat{m}_{iw|t} = \frac{1}{R}\sum_r \hat h_{iwr|t}$ and $\hat{\sigma}_{iw|t}^2 = \frac{1}{R-1}\sum_r (\hat h_{iwr|t}-\hat{m}_{iw|t})^2$. Then, the first order approximation of $\widehat{\text{cvDCL}}(t)^{INLA}$ as a function of the vector $\bm{h}_{|t}= \{h_{iwr|t}\}$ around the vector of means $\bm{m}_{|t}=\{m_{iw|t}\}$ is
\begin{equation} \label{eq:gdelta}
    \widehat{\text{cvDCL}}(t)^{INLA} = g(\bm{h}_{|t}) 
    \approx g(\bm{m}_{|t}) +  \sum_{i,w,r} (h_{iwr|t}-m_{iw|t})\left. \pdv{g}{h_{iwr|t}} \right\rvert_{ \bm{h}_{|t}=\bm{m}_{|t} } .
\end{equation}

Note that by construction $\mathrm{E}(g(\bm{h}_{|t})) \approx g(\bm{m}_{|t})$. Moreover, the partial derivative terms follow
\begin{equation*}
    \left. \pdv{g}{h_{iwr|t}} \right\rvert_{ \bm{h}_{|t}=\bm{m}_{|t} } = \frac{I(T_i>t)}{N_t}\frac{ \hat{p}(\bm{\theta}_w|\mathcal{D}) \Delta_w}{R  \sum_w \hat{p}(\bm{\theta}_w|\mathcal{D}) \Delta_w \hat{m}_{iw|t} }.
\end{equation*}

Additionally, given that the terms $h_{iwr|t}$ are independent and using the partial derivative expression from above, the variance of the expression in Equation \ref{eq:gdelta} follows 
\begin{equation} \label{eq:var_delta}
    \begin{split}
    \mathrm{Var}(\widehat{\text{cvDCL}}(t)^{INLA}) &\approx \sum_{i,w,r} \mathrm{Var}(h_{iwr|t}-m_{iw|t})\left(\left. \pdv{g}{h_{iwr|t}} \right\rvert_{ \bm{h}_{|t}=\bm{m}_{|t} } \right)^2\\
    &=R\sum_{i,w} \hat{\sigma}_{iw|t}^2 \left(\left. \pdv{g}{h_{iwr|t}} \right\rvert_{ \bm{h}_{|t}=\bm{m}_{|t} } \right)^2\\
    & = \frac{1}{N_t^2 R}\sum_{i}I(T_i>t)\frac{\sum_{w} \hat{\sigma}_{iw|t}^2 (\hat{p}(\bm{\theta}_w|\mathcal{D}) \Delta_w)^2}{(\sum_w \hat{p}(\bm{\theta}_w|\mathcal{D}) \Delta_w \hat{m}_{iw|t})^2}.
    \end{split}
\end{equation}

Thus, Equations \ref{eq:est_cvDCL_inla} and \ref{eq:var_delta} estimate $\text{cvDCL}(t)$ and its variance, respectively. 

Initially, \citet{rizopoulos2016personalized} suggested estimating $\text{cvDCL}(t)$ using posterior samples from an MCMC simulation, which is explained in detail in Appendix \ref{app:cvdcl_mcmc}. To assess the appropriateness of our estimate compared to the authors', we conducted a comparative analysis using simulated datasets described in Appendix \ref{app:sim_comp}.

\section{Empirical analysis on US mortgage prepayment} \label{sec:app}

\subsection{Data}\label{subsec:data}
We use the publicly available Single Family Loan-Level Dataset provided by Freddie Mac \footnote{Data can be accessed at \url{https://www.freddiemac.com/research/datasets/sf-loanlevel-dataset}}. This dataset contains comprehensive mortgage information, including loan-level details, application covariates, and monthly performance data.

\begin{figure}[ht]
    \centering
    \includegraphics[width=0.8\textwidth]{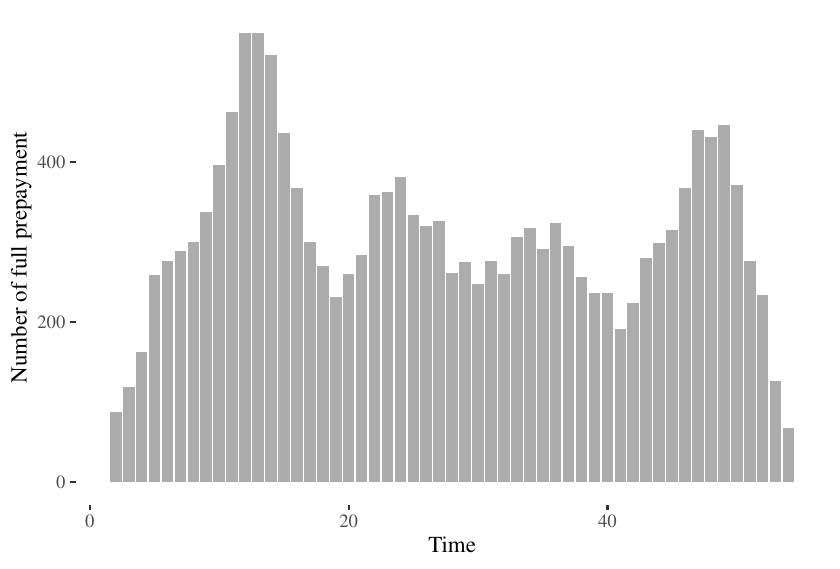}
    \captionof{figure}{Distribution of full prepayment events over time.}
    \label{fig:app_event} 
\end{figure}

The dataset consists of loans granted from June, 2015%
 to November, 2015%
, with monthly performance records tracked until December 2019. The mortgage with the longest time record is 4.5 years (54 months). The dataset considers 57,258%
 borrowers with a total of 2,559,056%
 observations. During the study period, 16,239%
 borrowers opted to prepay their mortgage loans fully. Figure \ref{fig:app_event} shows the distribution of full prepayment events over time.

The time-fixed covariates represented by the vector $\bm{z}_i$ in Equation \ref{eq:event} are as follows
\begin{itemize}[leftmargin=*]
    \item \textbf{cltv} is the loan-to-value ratio based on the original mortgage loan amount plus any other mortgage loan amount divided by the property's purchase price.
    \item \textbf{orig\_upb} is the original unpaid principal balance of the mortgage on the note date.
    \item \textbf{cnt\_units} denotes whether the mortgage is a one- ($=1$, 93\% of the loans) or more than one-unit property ($=0$, 7\% of the loans).
    \item \textbf{dti} is the debt-to-income ratio. It corresponds to the borrower's monthly debt payments divided by the total monthly income used to underwrite the loan.
    \item \textbf{int\_rt} is the interest rate given at the origination of the credit. 
    \item \textbf{term} corresponds to the number of scheduled monthly mortgage payments. It is divided between short-term loans, with terms less than or equal to 15 years ($=0$, 19\% of the loans) and long-term loans, with terms greater than 15 years ($=15$, 81\% the loans).
    \item \textbf{loan\_purpose} indicates whether the mortgage loan purpose is a cash-out refinance loan ($=0$, 22\% of the loans)\footnote{A cash-out refinance mortgage loan is a loan in which the use of the amount is not limited to specific purposes.}, no cash-out refinance loan ($=\text{N}$, 25\% of the loans) or purchase ($=\text{P}$, 53\% of the loans).
    \item \textbf{cnt\_borr} is the number of borrowers obligated to repay the mortgage. Either one borrower ($=0$, 48\% of the loans) or more than one ($=2$, 52\% of the loans).
\end{itemize}
These covariates are commonly used in similar contexts, as seen in studies such as \citet{wang2020reducing} and \citet{hu2019joint}, which also employ this dataset.

Table \ref{tab:desc_stats} shows descriptive statistics of the numeric covariates defined above. As a pre-processing step, these variables are standardised to have a zero-mean and standard deviation of 1.
\begin{table}[ht]
    \centering
    \footnotesize
    \begin{tabular}{l rrrr rrrr}
    \hline
    % latex table generated in R 4.1.0 by xtable 1.8-4 package
% Thu Dec 16 01:07:04 2021
$\text{Covariate}$ & $\text{N}$ & $\text{Mean}$ & $\text{SD}$ & $Q_{2.5\%}$ & $Q_{25\%}$ & $Q_{50\%}$ & $Q_{75\%}$ & $Q_{95\%}$ \\ 
  \hline
$\text{cltv}(\%)$ & 57258 & 73.50 & 16.97 & 38.00 & 65.00 & 79.00 & 85.00 & 95.00 \\ 
  $\text{orig\_upb}^{*}$ & 57258 & 256.32 & 121.87 & 88.00 & 161.00 & 241.00 & 336.00 & 475.00 \\ 
  $\text{dti}(\%)$ & 57258 & 34.87 & 9.14 & 19.00 & 28.00 & 36.00 & 42.00 & 48.00 \\ 
  $\text{int\_rt}(\%)$ & 57258 & 3.93 & 0.44 & 3.00 & 3.75 & 4.00 & 4.25 & 4.62 \\ 
   \hline
   
    \end{tabular} %}

    {\scriptsize \raggedright \textbf{*1,000 USD.} \par}
    \caption{Descriptive statistics for numeric covariates in the dataset.}
    \label{tab:desc_stats}
\end{table}

Concerning the longitudinal outcome, we are interested in a simple variable that indicates early repayments. \citet{medina2023joint_inla} show a correlation between borrowers who pay more than the due amount and the prepayment event for a dataset on consumer loans. For this empirical analysis, we follow the same rationale of looking for a candidate variable that measures the distance between what has been paid and what has been due. For simplicity, we look for a variable that shows a simple functional structure, such as a linear relationship, to facilitate the longitudinal design and, therefore, the model estimation.  

Assume that for a generic loan, we denote the interest rate given at origination as $i$ with a monthly instalment equal to $A$. Then, the sum of the total amount paid until time $t$, including the capitalisation of the inflows, is $A+A(1+i)+\ldots+A(1+i)^{t-1}$. Since the interest rate for mortgage loans is low for the analysed time period\footnote{The 95\% quantile of the annual interest rate is 4.62\%, which represents a monthly interest rate of 0.0038\%. See Table \ref{tab:desc_stats}.}, we can apply a first-order Maclaurin approximation (around zero) with respect to $i$, so we obtain $At + At(t-1)i/2 $. For the first periods, the linear term of this expression dominates. 

To make the longitudinal outcome comparable across different loans, we define it as $y_t =\sum_{s=0}^{t-1} (1+i)^s/T$. $T$ represents the study period, which is 54 months in our analysis. The sole purpose of including $T$ in the expression is for scaling. Note that by geometric series equivalence, the following expression also holds $y_t = \frac{(1+i)^t-1}{iT}$. If the total amount paid by the borrower is greater than the due amount at time $t$, then the observed $y_t$ should have a larger slope against time than the theoretical curve. 

In the dataset, we can access the observed unpaid principal balance at any given time. Since this amount may deviate from the scheduled balance, we aim to express $y_t$ in terms of this variable. Let's denote the unpaid principal balance at the loan's origination as $P_0$, the current unpaid principal balance at time $t$ as $P_t$, and the loan term as $M$. Therefore, using fundamental instalment relationships, we can derive an equivalent expression for $y_t$ as
\begin{equation*}
    y_t  = \frac{(P_0-P_t)}{P_0}\frac{(1+i)^{M}-1}{i T} .
\end{equation*}
Therefore, $y_t$ represents a longitudinal outcome with a simple structural form that can be expressed in terms of the observed flows.

\begin{figure}[ht]
    \centering
    \includegraphics[width=0.8\textwidth]{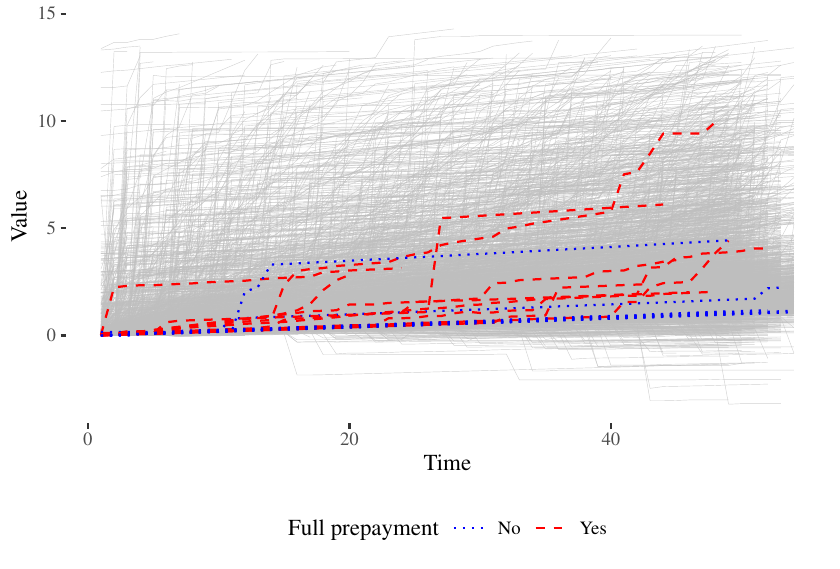}
    \captionof{figure}{Evolution of the longitudinal outcomes. Borrowers that experience prepayment are in dashed red line, and borrowers that are censored in blue dotted line.}
    \label{fig:app_long}
\end{figure}

Figure \ref{fig:app_long} shows our dataset's longitudinal outcome. To facilitate visualisation, mortgage loans that experienced prepayment are in a red dashed line, and those that do not are in a blue dotted line. 

\begin{figure}[ht]
    \centering
    \includegraphics[width=0.8\textwidth]{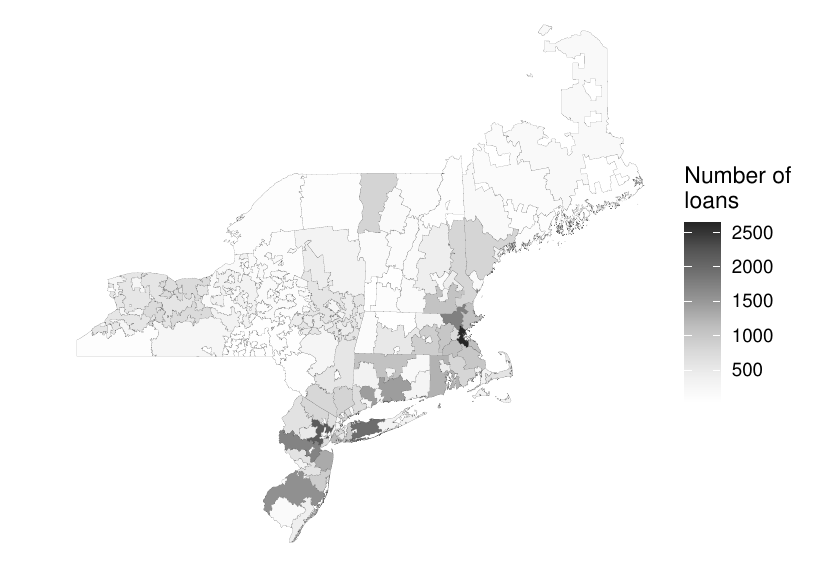}
    \captionof{figure}{Number of loans distributed by area.}
    \label{fig:app_map_n}
\end{figure}

Regarding geographical information, properties are located in 8 states: New York, New Jersey, Connecticut, Massachusetts, Rhode Island, Maine, New Hampshire and Vermont. These states are divided into 123 areas given by the first three digits of the zip code. The number of loans distributed among these areas is shown in the map of Figure \ref{fig:app_map_n}. 

\begin{figure}[ht]
    \centering
    \includegraphics[width=0.8\textwidth]{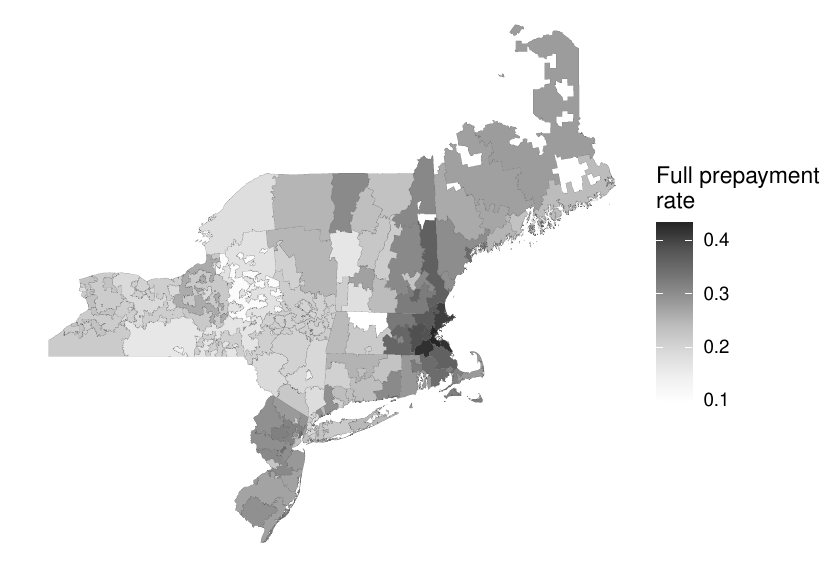}
    \captionof{figure}{Full prepayment rate distributed by area.}
    \label{fig:app_map_rt}
\end{figure}

In addition, Figure \ref{fig:app_map_rt} shows the corresponding full prepayment rates, calculated as the total number of events divided by the number of granted loans in each area. From this figure, although the rates include all events regardless of when they occurred, spatial clustering is observed and can be considered a first check to support the inclusion of spatial effects.

\subsection{Models and results} \label{subsec:mod}
Following the methodology described in Section \ref{subsec:est}, we estimate three joint models with different specifications. All these three specifications include the same time-fixed covariates described in Section \ref{subsec:data} and the same structure of the longitudinal outcome (see below). The differences come rather from the baseline specifications $\nu_{a,s}$ (see Equation \ref{eq:event}). Concretely, the longitudinal outcome follows
\begin{equation}\label{eq:app_long}
    \begin{split}
    Y_{i,s}|\eta_{Yi,s}, \tau_Y & \sim N(\eta_{Yi,s},\tau_Y^{-1})\\
    \eta_{Yi,s} &= \beta_{01} + \beta_{11} s + U_{0i} + U_{1i} s\\
    (U_{0i},U_{1i})^\intercal &\sim N_2(0, Q_{\bm{U}}^{-1}),
    \end{split}
\end{equation} 
where the covariance matrix $Q_{\bm{U}}^{-1}$ is parameterised via marginal precisions $\tau_{U_{0}}$ and $\tau_{U_{1}}$, and the pairwise correlation $\rho_{01}$ as follows
\begin{equation} \label{eq:pre_mat_stjm}
Q_{\bm{U}}^{-1} = \begin{pmatrix}
1/\tau_{U_{0}} & \rho_{01}/\sqrt{\tau_{U_{0}}\tau_{U_{1}}} \\
\rho_{01}/\sqrt{\tau_{U_{0}}\tau_{U_{1}}} & 1/\tau_{U_{1}}
\end{pmatrix}.
\end{equation}

The mixed-effect model from Equation \ref{eq:app_long} is known as intercept-slope random effects. This specification is justified because the longitudinal outcome approximates a linear trend when the interest rate is low, as shown in Section \ref{subsec:data}. Moreover, the survival process for the three models follow
\begin{equation}\label{eq:app_event}
    \begin{split}
    X_{i,s}|\eta_{Xi,s} &\sim \text{Bernoulli}(\logit^{-1}( \eta_{Xi,s})) \\
    \eta_{Xi,s} &= \nu_{a_i, s} + \bm{z}^{\intercal}_{i} \bm{\beta}_2 + \lambda (U_{0i} + U_{1i} s) ,
    \end{split}
\end{equation} 
where $\nu_{a_i, s}$ is the baseline risk. 

Table \ref{tab:models} describes the three models' specifications for $\nu_{a,s}$.  $M_1$ is a discrete-time joint model, similar to the ones analysed in \citet{medina2023joint_auto} (univariate, without autoregressive terms). $M_2$ introduces a novel aspect to the literature as there are no existing studies on joint models with spatial main effects in discrete time. The closest model to $M_2$ is a joint model that includes the spatial main effects in a Weibull survival component suggested by \citet{martins2016bayesian}. Finally, $M_3$ is the model that encompasses all the effects (the temporal and spatial main effects as well as the interactions).     

\begin{table}[ht]
\centering
\footnotesize
% \resizebox{0.9\textwidth}{!}{%
\begin{tabularx}{0.8\linewidth}{l c c c c} 
    \hline
    \textbf{Id} & \textbf{Temporal Effects} & \textbf{Spatial Effects} & \textbf{S-T Interactions} & $\nu_{a,s}$ \\
    \hline \hline
    $M_1$ & Yes & No & No & $\nu_0 + v_s$  \\
    \hline
    $M_2$ & Yes & Yes & No & $\nu_0 + v_s + u_a$  \\
    \hline
    $M_3$ & Yes & Yes & Yes & $\nu_0 + v_s + u_a + \delta_{a,s}$  \\
    \hline \\
\end{tabularx} %}
\caption{Specification of the joint models. $M_1$ includes the temporal effects in the baseline hazard. $M_2$ considers both the temporal and the spatial main effects, and $M_3$ captures both the temporal and the spatial main effects in addition to their interactions.} 
\label{tab:models}
\end{table}

Table \ref{tab:coef_all} shows the parameter estimates for the three models. We observe that the parameters strictly associated with the longitudinal outcome, $\beta_{01}$ and $\beta_{11}$, are consistent among $M_1$, $M_2$ and $M_3$. However, we notice differences in the covariates associated with the survival process. For instance, the coefficient related to \emph{cltv} under the estimation of $M_1$ is $0.301$, and its 95\% posterior credible interval does not include zero. The positive sign suggests that the higher the \emph{cltv}, the greater the probability of prepaying in full. Yet, when estimated under specifications $M_2$ and $M_3$, although the sign remains positive, the effect of this covariate decreases and is not as significant as before.  

In the same line, we notice that the effect of \emph{dti} for $M_1$ shows a negative relationship with the prepayment. Similar results were found in \citet{medina2023joint_inla} for the consumer loans dataset. However, when we include the spatial effects, either with $M_2$ or $M_3$, the relation of high \emph{dti} with a low probability of full prepayment is not entirely conclusive, even shifting the posterior marginals to the positive values.

For the other covariates, we found consistent results among the three models. For example, the original unpaid principal balance, \emph{orig\_upb}, shows a positive relationship with the prepayment, which is also supported by the prepayment models from \citet{medina2023joint_inla}. Moreover, it is less likely to prepay in full if the mortgage is more than a one-unit property (\emph{cnt\_units}), if the loan term is longer than 15 years (\emph{term\_g15}), if the number of borrowers is greater than 1 (\emph{cnt\_borr2}) or if the purpose of the loan is to purchase rather than refinance (\emph{loan\_purpose}). 

% \begin{landscape}
\begin{table}[ht]
\centering
\footnotesize
\begin{tabular}{l rrr rrr rrr}
\hline
& \multicolumn{3}{c}{$M_1$} &\multicolumn{3}{c}{$M_2$} &\multicolumn{3}{c}{$M_3$} \\
\cmidrule(r){2-4}  \cmidrule(r){5-7} \cmidrule(r){8-10} 
% latex table generated in R 4.2.0 by xtable 1.8-4 package
% Sat Jul  2 20:31:54 2022
 & $\text{Mean}$ & $2.5\%$ & $97.25\%$ & $\text{Mean}$ & $2.5\%$ & $97.25\%$ & $\text{Mean}$ & $2.5\%$ & $97.25\%$ \\ 
  \hline
$\beta_{01}$ & 0.011 & 0.008 & 0.013 & 0.011 & 0.008 & 0.013 & 0.011 & 0.008 & 0.013 \\ 
  $\beta_{11}$ & 0.027 & 0.026 & 0.027 & 0.027 & 0.026 & 0.027 & 0.027 & 0.026 & 0.027 \\ 
  $\nu_0$ & -8.247 & -8.451 & -8.043 & -8.590 & -8.798 & -8.382 & -8.541 & -8.748 & -8.334 \\ 
  $\text{cltv}$ & 0.301 & 0.081 & 0.520 & 0.114 & -0.120 & 0.347 & 0.117 & -0.115 & 0.349 \\ 
  $\text{orig\_upb}$ & 0.143 & 0.128 & 0.159 & 0.153 & 0.134 & 0.172 & 0.153 & 0.134 & 0.172 \\ 
  $\text{cnt\_units1}$ & 0.439 & 0.367 & 0.510 & 0.425 & 0.351 & 0.499 & 0.408 & 0.334 & 0.482 \\ 
  $\text{dti}$ & -0.109 & -0.222 & 0.005 & 0.024 & -0.091 & 0.139 & 0.018 & -0.097 & 0.132 \\ 
  $\text{int\_rt}$ & 0.794 & 0.743 & 0.845 & 0.869 & 0.817 & 0.921 & 0.846 & 0.794 & 0.898 \\ 
  $\text{term\_g15}$ & -0.458 & -0.518 & -0.399 & -0.531 & -0.591 & -0.470 & -0.518 & -0.579 & -0.458 \\ 
  $\text{loan\_purposeN}$ & 0.028 & -0.018 & 0.074 & -0.005 & -0.051 & 0.041 & -0.010 & -0.056 & 0.036 \\ 
  $\text{loan\_purposeP}$ & -0.137 & -0.180 & -0.093 & -0.100 & -0.144 & -0.057 & -0.109 & -0.153 & -0.066 \\ 
  $\text{cnt\_borr2}$ & -0.060 & -0.091 & -0.028 & -0.061 & -0.093 & -0.029 & -0.063 & -0.095 & -0.031 \\ 
  $\lambda$ & 0.201 & 0.188 & 0.211 & 0.146 & 0.139 & 0.156 & 0.171 & 0.162 & 0.184 \\ 
  $\tau_Y$ & 34.713 & 34.653 & 34.777 & 34.715 & 34.656 & 34.783 & 34.784 & 34.730 & 34.826 \\ 
  $\tau_{U_{0}}$ & 9.651 & 9.568 & 9.752 & 9.627 & 9.557 & 9.716 & 9.870 & 9.767 & 9.956 \\ 
  $\tau_{U_{1}}$ & 895.736 & 885.777 & 904.230 & 887.034 & 874.223 & 898.767 & 880.029 & 871.477 & 890.400 \\ 
  $\rho_{01}$ & -0.067 & -0.076 & -0.058 & -0.052 & -0.060 & -0.046 & -0.056 & -0.064 & -0.051 \\ 
  $\tau_v$ & 3.004 & 2.245 & 3.653 & 2.435 & 1.369 & 3.375 & 0.802 & 0.518 & 1.094 \\ 
  $\tau_u$ &  &  &  & 16.667 & 12.163 & 25.503 & 14.777 & 12.806 & 17.464 \\ 
  $\tau_{\delta}$ &  &  &  &  &  &  & 104.613 & 57.088 & 196.704 \\ 
   \hline
 
\end{tabular} %}
\caption{Parameter estimates for models $M_1$, $M_2$ and $M_3$.}
\label{tab:coef_all}
\end{table}
% \end{landscape}

The interest rate granted at origination, \emph{int\_rt}, is expected to play an essential role in the decision of full prepayment. If the reference interest rates fall compared to the one granted, it is more attractive to renegotiate the credit. As seen in Table \ref{tab:coef_all} for all three models, its positive effect suggests that the higher the interest rate given at origination, the greater the probability of full prepayment. This has also been seen in \citet{medina2023joint_inla}. However, when we include the spatial effects, we note that the associated coefficient also increases.

Regarding $\lambda$, the parameter that associates the random effects of the longitudinal outcome and the survival process, we observe that the three models estimate a significant positive association, as expected, since the more is paid off from what is owed, the more likely it is to prepay in full. However, the magnitude of the estimate differs among the models. The largest one is due to $M_1$ with a mean of $0.201$. When we add the spatial main effects in $M_2$, we see a mean decrease to $0.146$. Yet, when we add the spatio-temporal interactions in $M_3$, we observe a value in between, with a mean of $0.171$. 

\begin{figure}[ht]
    \centering
      \includegraphics[width=.8\textwidth]{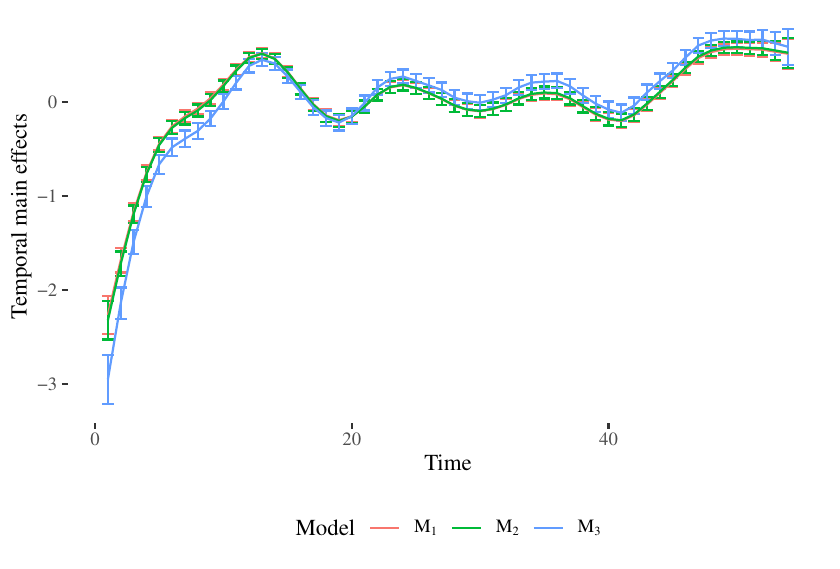}
      \caption{Temporal main effects estimated by the three models. The error bars represent the estimated 95\% credible intervals.}
      \label{fig:temp_main_eff}
\end{figure}

Furthermore, we obtained similar results among the three models regarding the hyperparameters associated with the longitudinal outcome. Namely, the precision of the error terms $\tau_Y$ and the elements $\tau_{U_0}$, $\tau_{U_1}$ and $\rho_{01}$ of the precision matrix $Q_{\bm{U}}$. However, the precision of the temporal main effects $\tau_v$ changes among the three models. We see a mean of $3.004$, $2.435$ and $0.802$ for models $M_1$, $M_2$ and $M_3$, respectively. That raises the question of how different each model's estimated temporal main effects are. Figure \ref{fig:temp_main_eff} shows the estimated temporal main effects for the three models. Models $M_1$ and $M_2$ overlap for much of the study period, and $M_3$ shows some differences, in particular for the first periods, but overall, the effect of the three models is fairly comparable. 

To compare the performance of the models, we follow the procedure described in Section \ref{subsec:inla_cvdcl}. We estimate the $\widehat{\text{cvDCL}}(t)^{INLA}$ (Equation \ref{eq:est_cvDCL_inla}) for six evaluation times $t$, ranging from 12 to 42 months with an increment of 6 months. The results are shown in Table \ref{tab:res_cvdcl} (we deliberately omit the word \say{INLA} to shorten the notation). $N_t$ is the number of borrowers at risk, and the values in brackets are the estimates of Monte Carlo standard error derived from Equation \ref{eq:var_delta}. It is worth noting that the metric value should be compared across the models for one value of $t$, that is, all the values that belong to the same row since, between the rows, there is an evident overlap of datasets. The table shows that both $M_2$ and $M_3$ outperform $M_1$. Adding the latent spatial component can increase the model's performance for this dataset. However, when we compare models $M_2$ and $M_3$, that is, when we add on top of the spatial main effects, the spatio-temporal interactions, the improvements are not as clear as before. 
\begin{table}[ht]
\centering
\resizebox{0.8\textwidth}{!}{%
\sisetup{detect-weight,mode=text}
\renewrobustcmd{\bfseries}{\fontseries{b}\selectfont}
\renewrobustcmd{\boldmath}{}
\newrobustcmd{\B}{\bfseries}
\begin{tabular}{l cccc}
\hline
% latex table generated in R 4.1.0 by xtable 1.8-4 package
% Fri Dec 24 15:48:35 2021
 & $N_t$ & $M_1$ & $M_2$ & $M_3$ \\ 
  \hline
$\widehat{\text{cvDCL}}(t=12)$ & 53963 & 1.4438 (5.69e-06) & 1.4244 (1.03e-05) & 1.4304 (5.72e-05) \\ 
  $\widehat{\text{cvDCL}}(t=18)$ & 51489 & 1.2231 (4.01e-06) & 1.2143 (7.57e-06) & 1.2165 (2.20e-05) \\ 
  $\widehat{\text{cvDCL}}(t=24)$ & 49607 & 1.0349 (3.58e-06) & 1.0306 (6.91e-06) & 1.0310 (1.35e-05) \\ 
  $\widehat{\text{cvDCL}}(t=30)$ & 47839 & 0.8472 (3.27e-06) & 0.8450 (6.40e-06) & 0.8448 (1.15e-05) \\ 
  $\widehat{\text{cvDCL}}(t=36)$ & 46059 & 0.6453 (2.91e-06) & 0.6438 (5.68e-06) & 0.6439 (1.08e-05) \\ 
  $\widehat{\text{cvDCL}}(t=42)$ & 44611 & 0.4656 (2.52e-06) & 0.4644 (4.96e-06) & 0.4648 (1.00e-05) \\ 
   \hline
 
\end{tabular}}
\caption{Comparison of model performance. The value in brackets is an estimate of the Monte Carlo standard error.}
\label{tab:res_cvdcl}
\end{table}

\begin{figure}[ht]
    \centering
    \includegraphics[width=1\textwidth]{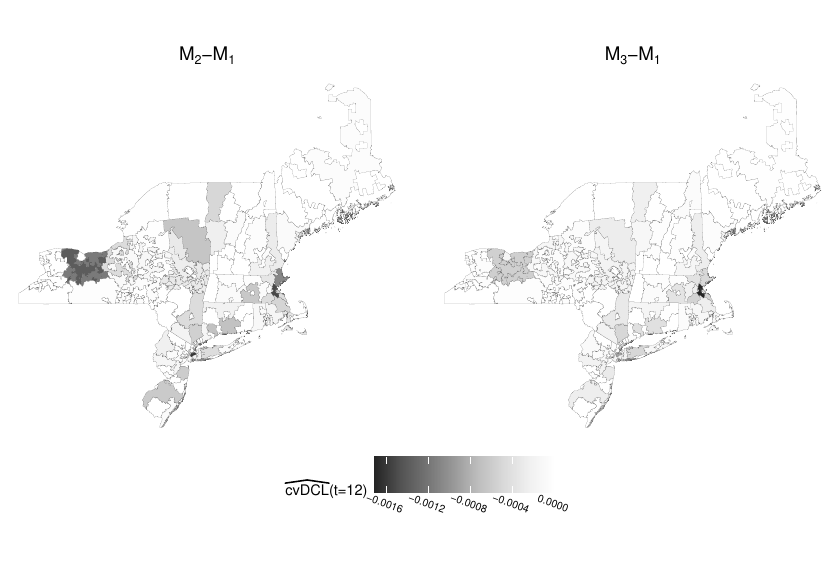}
    \captionof{figure}{Difference between the $\widehat{\text{cvDCL}}(t=12)$ for models $M_2$ and $M_3$ with respect to $M_1$ and segmented by area.}
    \label{fig:app_map_cvDCL_12}
\end{figure}

To further explore the analysis, we assign the overall $\widehat{\text{cvDCL}}(t)$ to the corresponding area and compare from which areas the major gains are obtained for models $M_2$ and $M_3$ with respect to model $M_1$. Figure \ref{fig:app_map_cvDCL_12} shows two maps for the segmented $\widehat{\text{cvDCL}}$ evaluated at $t=12$. The left corresponds to the difference between $M_2$ and $M_1$ ($M_2-M_1$), and the right one to $M_3-M_1$. From both maps, we observe that the major contributions to the overall metric mainly come from the middle-left (west, Rochester area) and middle-right parts (east, Boston area) of the maps. These differences are increased for model $M_2$. 

\begin{figure}[ht]
    \centering
    \includegraphics[width=1\textwidth]{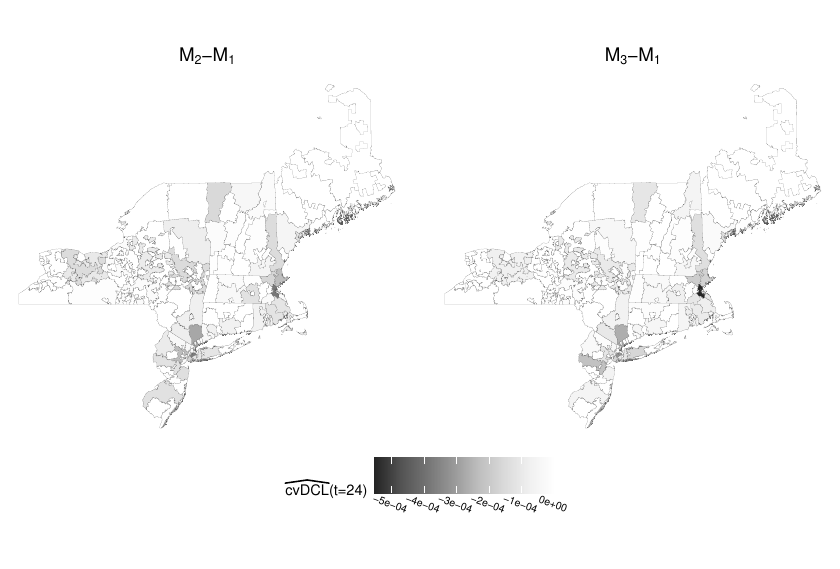}
    \captionof{figure}{Difference between the $\widehat{\text{cvDCL}}(t=24)$ for models $M_2$ and $M_3$ with respect to $M_1$ and segmented by area.}
    \label{fig:app_map_cvDCL_24}
\end{figure}

Moreover, when we choose a different evaluation time, for instance, $t=24$ (see Figure \ref{fig:app_map_cvDCL_24}), now the contributions coming from the Rochester area are not as meaningful as for $t=12$. Instead, the differences come from New Jersey, New York City and Boston locations. Thus, when we include spatial effects, we see that the consistent improvements in the performance evaluated in different periods are not exclusively attributed to a particular area.

\section{Discussion} \label{sec:con}
Previous studies have shown that the joint model approach offers advantages over the widely used survival approaches in credit-related applications \citep{hu2019joint,medina2023joint_auto,medina2023joint_inla}. In this manuscript, we extend this existing research by investigating enhanced representations of the survival predictor. Specifically, we incorporate spatial and spatio-temporal effects into the baseline hazard and explore how this modification can impact the prediction performance of a prepayment model for US mortgages. This decision is supported by two main factors. Firstly, there is evidence from previous research that incorporating spatial effects into credit risk models can result in improved predictions \citep{calabrese2020spatial,medina2022spatial}. Secondly, as mentioned earlier, the joint modelling of longitudinal and survival processes offers an attractive dynamic prediction framework for credit modelling.

In this respect, we make four main contributions. First, we introduce the Spatio-Temporal Joint Model (STJM), a Bayesian joint model formulated in discrete time that includes a flexible baseline hazard in the survival predictor. The baseline hazard is effectively decomposed into temporal and spatial main effects, along with their interactions. For this latter, we leverage the approach from \citet{clayton1996generalized} in which the structure matrix is built by the Kronecker product of the main effects structure matrices. Moreover, we follow the \citet{goicoa2018spatio} approach to get appropriate identifiability constraints by using spectral decomposition over the structure matrices. 

Second, to estimate the STJM in a large dataset, we formulate the model using the INLA methodology \citep{rue2009approximate} and implement it in the \texttt{R-INLA} package (\url{https://www.r-inla.org/}). This implementation allows us to scale the model to a dataset with 57,258%
 borrowers with 2,559,056%
 total observations. As far as we know, this is the largest sample size used in a joint model application. 

Third, we propose a modified version of the \emph{cross-validated Dynamic Conditional Likelihood} originally proposed by \citet{rizopoulos2016personalized}. Our adaptation leverages the estimations obtained through the INLA methodology, which differs from the original version that relies on posterior MCMC samples, resulting in reduced computational costs. We compare the original and the proposed versions by a simulation study that demonstrates adequate results (see Appendix \ref{app:sim_comp}).

Fourth, we apply the proposed approach to predict the full prepayment event in US mortgage loans. The analysis comprises three models: (1) measuring only the temporal main effect ($M_1$), (2) adding the temporal and spatial main effects ($M_2$), and (3) incorporating both main effects along with their interactions ($M_3$). In general, the parameter estimates show agreement across the three joint models. However, a notable difference arises concerning the covariate \say{debt to income ratio} (\emph{dti}), which represents the sum of the borrower's monthly debt payments divided by their monthly income. When spatial effects are excluded, the parameter estimate indicates that higher \emph{dti} values are associated with a lower probability of prepayment. However, this relationship no longer holds when spatial effects are included.

Additionally, the empirical results reveal that spatial effects consistently enhance the prediction performance of the joint model across all evaluation times. Interestingly, when we contrast the performance evaluated at different time intervals, these improvements are not limited to a specific region.  However, including spatio-temporal interactions does not yield equally clear performance gains compared to the model without such interactions. 

Our robust findings, facilitated by the suitability of INLA estimation for large datasets, suggest that mortgage grantors could improve predictive performance in practice by employing the proposed spatial approach. Potential impacts may involve the refinement of methodologies for estimating cash flows for credit loss provisioning purposes and the identification of optimal levels of economic capital, ultimately leading to a more competitive and prudent risk assessment.

This study presents compelling insights that undoubtedly pave the way for further research. One avenue worth exploring is the inclusion of TVCs whose processes do not need to be jointly estimated with a dependent variable. These TVCs could encompass macroeconomic variables, allowing us to investigate how changes in the overall economic conditions impact the joint model's performance. Comparable approaches have been adopted in credit survival models \citep{bellotti2009credit,djeundje2018incorporating,dirick2019macro}, corporate credit default models \citep{li2022corporate} and a time-continuous joint model without spatial effects \citep{hu2019joint}. Since these TVCs are external factors, we can assume that the occurrence of a specific event does not influence their trajectories. Consequently, there is no need for a borrower-specific longitudinal model for these covariates. This could lead to a more comprehensive joint model framework, leveraging individual predictions while incorporating the influence of economic conditions.

\bibliographystyle{apalike}  
\bibliography{references} 

\appendix

\section{Estimation of cvDCL under MCMC scheme} \label{app:cvdcl_mcmc}
In Section \ref{subsec:inla_cvdcl}, we show how the \emph{cross-validated Dynamic Conditional Likelihood} (cvDCL) is estimated using the INLA methodology. Here, we describe how the cvDCL is computed with an MCMC sampling scheme. This is done for comprehensive understanding since in Appendix \ref{app:sim_comp}, we compare numerically how different these two approaches are using simulation analysis.

From Equation \ref{eq:cvDCL}, we know
\begin{equation*} 
    \text{cvDCL}(t) = \frac{1}{N_t}\sum_{i = 1}^N-I(T_i>t)\log\{p(T_i,\delta_i | T_i>t, \bm{y}_i(t),\mathcal{D}_{-i} ) \},
\end{equation*}
where $N_t = \sum_{i=1}^N I(T_i>t)$ (the number of loans at risk at time $t$). It can be shown that \citep[see][for further details]{rizopoulos2016personalized}\footnote{In that work, Equation \ref{eq:inv_prob2} is presented as equality. We confirmed with the author that there is an error and that it should be an approximation symbol instead.}
\begin{equation}\label{eq:inv_prob2}
        p(T_i,\delta_i | T_i>t, \bm{y}_i(t),\mathcal{D}_{-i} )^{-1} \approx \int \frac{p(\bm{U}_i,\bm{\Theta}|\mathcal{D})}{p(T_i,\delta_i | T_i>t, \bm{y}_i(t),\bm{U}_i,\bm{\Theta} )}  \mathrm{d}\bm{\Theta} \mathrm{d}\bm{U}_i,
\end{equation}
where $\bm{\Theta}$ is the set of all parameters as described in Section \ref{subsec:est} and $\bm{U}_i$ the random effects for loan $i$. Let $\bm{\Theta}^{(g)}$ and $\bm{U}_i ^{(g)}$ denote the $g$-th realisation of the posterior sample with $g = 1,...,G$, then $p(T_i,\delta_i | T_i>t, \bm{y}_i(t),\mathcal{D}_{-i} )^{-1}$ can be estimated by
\begin{equation*} 
        \hat p(T_i,\delta_i | T_i>t, \bm{y}_i(t),\mathcal{D}_{-i} )^{-1} = \frac{1}{G}\sum_{g=1}^G \frac{1}{p(T_i,\delta_i | T_i>t, \bm{y}_i(t),\bm{U}_i ^{(g)},\bm{\Theta}^{(g)} )}.
\end{equation*}
Hence, $\text{cvDCL}(t)$ can be computed as
\begin{equation} \label{eq:est_cvDCL_mcmc}
    \widehat{\text{cvDCL}}(t)^{MCMC} = \frac{1}{N_t}\sum_{i = 1}^N I(T_i>t)\log\left\{\frac{1}{G}\sum_{g=1}^G \frac{1}{p(T_i,\delta_i | T_i>t, \bm{y}_i(t),\bm{U}_i ^{(g)},\bm{\Theta}^{(g)} )} \right\}.
\end{equation}
We estimate the variance of $\widehat{\text{cvDCL}}(t)^{MCMC}$ through batching \citep{carlin2000bayes}. This requires that a long run of $G$ samples is divided into $M$ successive batches of size $H$ (i.e.\ $G=M \cdot H$). For each batch $m = 1,...,M$, we calculate $\widehat{\text{cvDCL}}(t)^{MCMC}_m$ using its $H$ posterior samples, and the variance is then the sample variance of these $M$ estimations.  

\section{Comparison cvDCL: MCMC and INLA} \label{app:sim_comp}
Here, we study how different is the estimation of the cvDCL between the MCMC and INLA procedures (Equations \ref{eq:est_cvDCL_mcmc} from Appendix \ref{app:cvdcl_mcmc} and \ref{eq:est_cvDCL_inla} from Section \ref{subsec:inla_cvdcl}, respectively). To this end, we first generate data from a joint model that follows Equations \ref{eq:cvsim_long} and \ref{eq:cvsim_event} for the longitudinal and event processes, respectively.  
\begin{equation}\label{eq:cvsim_long}
\begin{split}
  Y_{i,s}|\eta_{Yi,s}, \tau_Y & \sim N(\eta_{Yi,s},\tau_Y^{-1})\\
    \eta_{Yi,s} &= \beta_{01} + U_{0i} + (\beta_{11} + U_{1i})s ,\\
     (U_{0i}, U_{1i})^\intercal &\sim N_{2}(\mathbf{0}, Q^{-1}_{\mathbf{U}}),
\end{split}
\end{equation}
\begin{equation}\label{eq:cvsim_event}
 \begin{split}
   X_{i,s}|\eta_{Xi,s} &\sim \text{Bernoulli}(\logit^{-1}( \eta_{Xi,s})) \\
    \eta_{Xi,s} &= \nu_0 + v_s + \beta_{12} z_{1i} + \beta_{22} z_{2i} + \lambda (U_{0i} + U_{1i} s) ,\\
    v_s &\sim RW2(\tau_v).
\end{split}
\end{equation}
Next, we estimate the $\widehat{\text{cvDCL}}(t)^{MCMC}$ and $\widehat{\text{cvDCL}}(t)^{INLA}$, for different values of $t$, assuming two different specifications of the joint model. The first specification is the correct one, i.e.\ follows Equations \ref{eq:cvsim_long} and \ref{eq:cvsim_event}. The second one omits the second covariate in the event predictor. Specifically, it assumes the linear predictor of the event process as $\nu_0 + v_s + \beta_{11} z_{1i} + \lambda (U_{0i} + U_{1i} s)$ (see Equation \ref{eq:cvsim_event}). By comparing the cvDCL values under these two distinct settings, we not only gain insights into their differences but also assess how the cvDCL varies when one specification outperforms the other in explaining the data.

Table \ref{tab:sim_perf} shows the results of the comparative analysis. The INLA implementation is presented in two ways, one calculates $h_i(\bm{\theta}_w,\bm{\mu}_{-\bm{U}_i}^{(r,w)} | t)$ (see Section \ref{subsec:inla_cvdcl}) with the Laplace method (INLA Lap) and the other with empirical Bayes (INLA EB).

\begin{table}[ht]
\centering
\footnotesize
% \resizebox{\textwidth}{!}{%
\begin{tabular}{l r ccc}
\hline
& & \multicolumn{3}{c}{Correct Specification}  \\
\cmidrule(r){3-5} 
% latex table generated in R 4.2.0 by xtable 1.8-4 package
% Sat Jul  2 20:53:47 2022
 & $N_t$ & $\text{MCMC}$ & $\text{INLA Lap}$ & $\text{INLA EB}$ \\ 
  \hline
$\widehat{\text{cvDCL}}(t=12)$ & 424 & 2.5915 (6.80e-04) & 2.5922 & 2.5778 \\ 
  $\widehat{\text{cvDCL}}(t=18)$ & 347 & 2.4715 (6.75e-04) & 2.4716 & 2.4679 \\ 
  $\widehat{\text{cvDCL}}(t=24)$ & 183 & 2.3951 (8.66e-04) & 2.3942 & 2.3930 \\ 
  $\widehat{\text{cvDCL}}(t=30)$ &  85 & 2.1884 (1.63e-03) & 2.1857 & 2.1864 \\ 
  $\widehat{\text{cvDCL}}(t=36)$ &  36 & 1.7153 (4.22e-03) & 1.7085 & 1.7115 \\ 
   \hline

& & \multicolumn{3}{c}{Other Specification} \\
\cmidrule(r){3-5} 
% latex table generated in R 4.2.0 by xtable 1.8-4 package
% Sat Jul  2 20:53:48 2022
 & $N_t$ & $\text{MCMC}$ & $\text{INLA Lap}$ & $\text{INLA EB}$ \\ 
  \hline
$\widehat{\text{cvDCL}}(t=12)$ & 424 & 2.8023 (7.34e-04) & 2.8027 & 2.7949 \\ 
  $\widehat{\text{cvDCL}}(t=18)$ & 347 & 2.6893 (7.95e-04) & 2.6890 & 2.6874 \\ 
  $\widehat{\text{cvDCL}}(t=24)$ & 183 & 2.6037 (1.27e-03) & 2.6027 & 2.6031 \\ 
  $\widehat{\text{cvDCL}}(t=30)$ &  85 & 2.3634 (1.70e-03) & 2.3598 & 2.3632 \\ 
  $\widehat{\text{cvDCL}}(t=36)$ &  36 & 1.8390 (4.27e-03) & 1.8305 & 1.8382 \\ 
   \hline

\end{tabular} %}
\caption{Comparison of model performance for simulated data and two different specifications.}
\label{tab:sim_perf}
\end{table}

\end{document}